\documentclass{aa}
\usepackage{graphics}

\begin{document}
\thesaurus{11(11.03.1;11.03.4 Abell~2104;11.04.1;11.05.1;11.06.2),12(12.04.1;12.07.1)}

\title{Probing the gravitational potential of a nearby lensing cluster \object{Abell~2104}}
\subtitle{}

\author{H. Liang\inst{1} \and L. L\'emonon\inst{2} \and I. Valtchanov\inst{2} \and M. Pierre\inst{2} \and G. Soucail\inst{3} }

\offprints{H. Liang}

\institute{Department of Physics, University of Bristol, Tyndall Avenue., Bristol BS8 1TL, UK\\
email: h.liang@bristol.ac.uk
\and
CEA Saclay DSM/DAPNIA/SAp, Service d'Astrophysique, F-91191
Gif sur Yvette, France\\
\and
Observatoire Midi Pyr\'en\'ees, Laboratoire d'Astrophysique de Toulouse, URA 285, 14 Avenue E. Belin, F-31400 Toulouse, France}

\date{}  
\titlerunning{Probing the gravitational potential of Abell~2104}
\maketitle

\begin{abstract}
The cluster \object{Abell~2104} is one of the lowest redshift clusters
($z=0.153$) known to have a gravitational lensing arc. We present
detailed analysis of the cluster properties such as the gravitational
potential using the X-ray data from {\it ROSAT} (HRI) and {\it ASCA}, as well as
optical imaging and spectroscopic data from the CFHT. The cluster is
highly luminous in the X-ray with a bolometric luminosity of
$L_{x}\sim 3\times 10^{45}$ ergs~s$^{-1}$ and a high gas temperature
of $\sim 10.4$ keV. The X-ray emission extending out to at least a
radius of 1.46 Mpc, displays significant substructure.  The total mass
deduced from the X-ray data under the assumption of hydrostatic
equilibrium and isothermal gas, is found to be
$M_{tot}(r<1.46\mathrm{Mpc})\sim (8.0\pm 0.8)\times 10^{14}
M_{\sun}$. The gas fraction within a radius of 1.46 Mpc is $\sim
5-10$\%. The cluster galaxy velocity distribution has a dispersion of
$1200\pm200$ km\,s$^{-1}$ with no obvious evidence for
substructure. The total mass within 1.46 Mpc, deduced from Jean's
equation using the observed galaxy number density distribution and
velocity dispersion, is found to be $\sim 6.8\times 10^{14} M_{\sun}$
to $\sim 2.6\times 10^{15} M_{\sun}$ marginally consistent with the
X-ray deduced total mass.

\keywords{galaxies: clustering -- clusters of galaxies: individual
\object{Abell~2104} -- cosmology: observations -- dark matter}
\end{abstract}

\section {Introduction}

Clusters of galaxies are the largest bound systems in the Universe,
and as such they are the largest objects where detailed studies of
their gravitational potential are possible. Given their large sizes, 3
to 6 Mpc in extent, they are also thought to be representative of the
Universe in terms of the baryonic fraction which is directly related
to the density of the universe and the predictions of the Big Bang
nucleosynthesis theory. Studies so far have found that the baryonic
fractions in clusters favour a low matter density universe given the
predictions of baryon densities given by the nucleosynthesis theory
(e.g. White et al. 1993). Recently, detailed and {\it independent}
estimates of cluster total mass distributions have become available; the
mass--tracers used and the observational techniques employed can be
summarised as follows:

\begin{itemize}
\item {\it Cluster Galaxies} : these have a long tradition of providing
mass estimates via application of the Virial Theorem to the observed 
dispersion in their radial velocities. The method rests upon the assumption
that the galaxies are in dynamical equilibrium.
\item {\it Hot Intracluster Gas} : as well as being an important mass
component of clusters, its X-ray emission provides an ideal tracer --
through the hydrostatic equation -- of the total underlying mass. The
assumption that the gas is in hydrostatic equilibrium with the
cluster's gravitational potential is thought to be reasonably secure
for the central few Mpc (Evrard et al. 1996 and Schindler 1996)
and the gas density and temperature profiles required to solve the
hydrostatic equation are readily available from the X-ray data.
\item {\it Gravitational Lensing} : here the lensing action of the
cluster on background sources, as revealed in deep high resolution
imagery (Tyson et al. 1990; Fort \& Mellier 1994 and references there
in), is used to provide a direct measure of the shape and depth of the
cluster potential and hence the projected mass distribution (Kaiser \&
Squires 1993, Broadhurst et al. 1995 etc.). Unlike the first 2
methods, this approach is not reliant upon assumptions of hydrostatic
or dynamical equilibrium.
\end{itemize}

For detailed studies in the X-ray and optical, we need a nearby
cluster, though gravitational lensing effects are diminished for low
redshift clusters.  An ideal cluster for this kind of detailed and
independent estimates of mass distributions, would be one of the
lowest redshift clusters with obvious lensing effects such as a giant
arc.  In this paper, we will analyse the X-ray and optical data for
one of the nearby lensing clusters.  

\object{Abell~2104} is a rich cluster (richness class 2) at a redshift
of 0.155 (Allen et al. 1992). It was found to have a high X-ray
luminosity from the {\it ROSAT} all-sky survey data (Pierre et al.
1994). Subsequent optical followup observations with the CFHT revealed
an arc embedded in the halo of the central cD galaxy $7\farcs2$ away
from the centre (Pierre {\em et al.}  1994).  The arc spans
$10\arcsec$ in length and it is amongst the reddest known
arcs. Fig.~\ref{f:arc} shows a close up picture of the arc.  Given the
small arc radius, it is important to have a high resolution X-ray
observation with an instrument such as the {\it ROSAT}/HRI to probe the
gravitational potential within the arc radius.

The optical data including photometry and spectroscopy will be
analysed in Sec. 2. The spatial and spectroscopic analysis of the
X-ray data from {\it ROSAT} and {\it ASCA} will be given in Sec. 3. The
independent mass estimates using different methods as well as a
comparisons will be given Sec. 4.

Throughout the paper we adopt a cosmological model with $H_0=50$
km\,s$^{-1}$Mpc$^{-1}$, $\Omega_{0} = 1$ and $\Lambda_{0}=0$. Celestial
coordinates are in J2000.

\section{Optical data}

\subsection{Observations}

\begin{figure*}
\resizebox{\hsize}{!}{\includegraphics{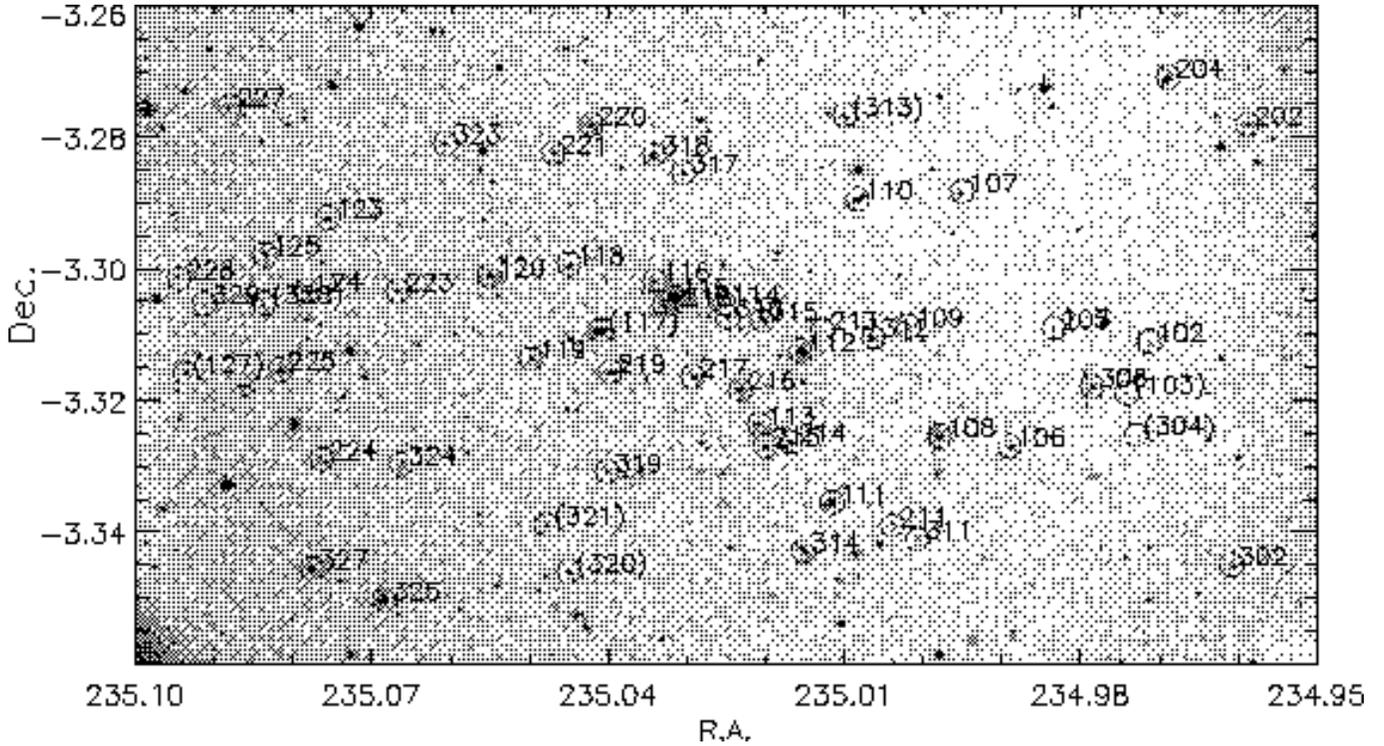}}
\caption []{ Finding chart for galaxies with measured redshifts.
 The reference numbers are the
same as in Table~\ref{spectro}. The non-member galaxies are 
marked with a bracket around their reference number.
\label{chart} }
\end{figure*}

The data were collected in 4 nights at the 3.6 m CFHT Telescope in May
1993. Two 10 minutes exposures in B band and two 15 minutes exposures
in R band were obtained. Exposures of 30 to 55 minutes per
spectroscopic mask was obtained for 3 separate masks, each containing
about 30 slits (Fig.~\ref{chart}).  The focal reducer MOS/SIS together
with CCD Lick2 ($2048 \times 2048$ pixels of 15 $\mu$m) were used
during the run. This CCD is a thick device having a quantum efficiency
of $\sim 10\%$ in the blue. The observing configuration provides a
pixel size of $0.314\arcsec$ over a field of view of about $10\arcmin
\times 10\arcmin$. The overall image quality was good (stellar FWHM
$\sim 0.9\arcsec $) although some optical distortions were conspicuous
near the edges of the images due to the optics of the focal reducer.

\begin{figure*}
\resizebox{\hsize}{!}{\includegraphics{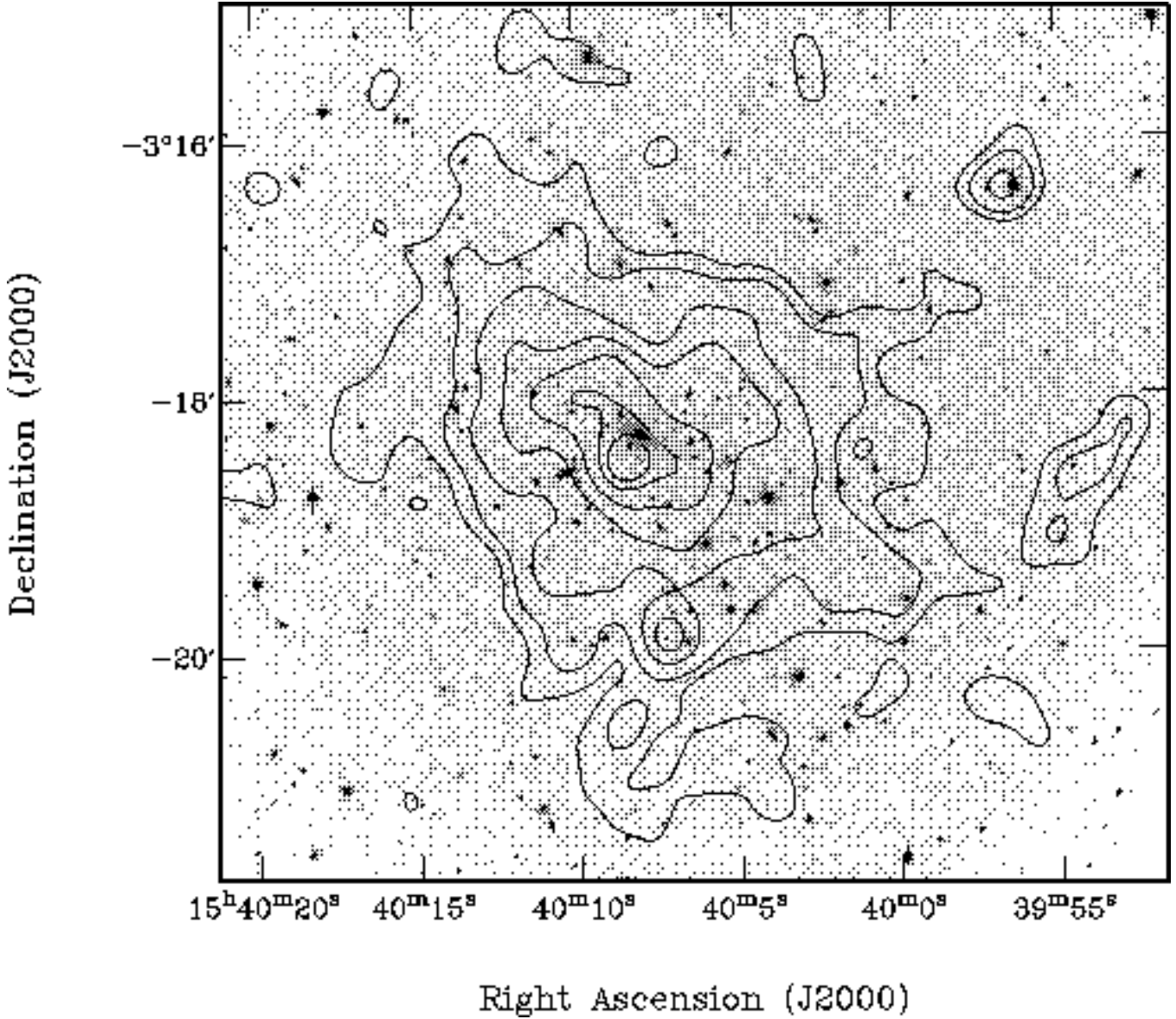}}
\caption []{ Optical field of \object{Abell~2104}, observed at the
CFHT in R band.  Overlaid are the {\it ROSAT} HRI contours with levels
$(1.7, 2.0, 2.6, 3.2, 3.8, 4.3, 4.6)\times
10^{-6}$~counts\,\,s$^{-1}$\,arcsec$^{-2}$. The X-ray image was
rebinned into $2^{''}$ pixels and smoothed with a $10^{''}$ Gaussian.
\label{superover} }
\end{figure*}

\begin{figure}
\resizebox{\hsize}{!}{\includegraphics{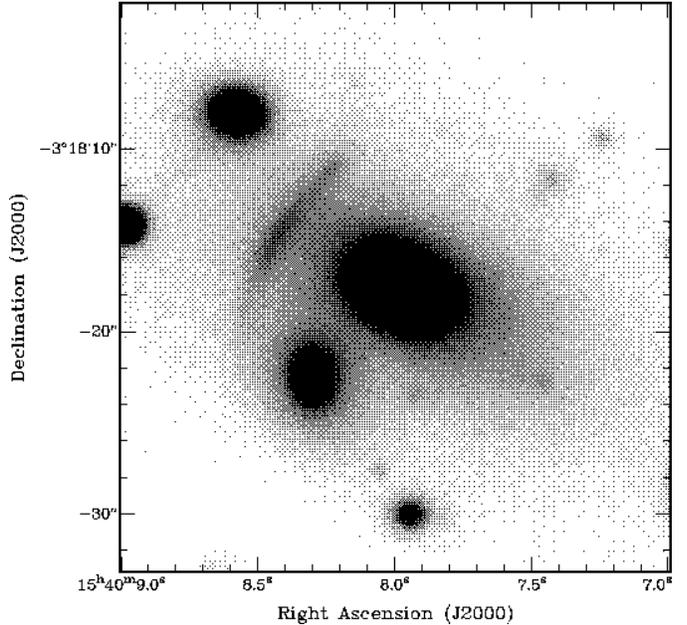}}
\caption{A close up image of the central regions of \object{Abell~2104} showing the giant arc.}
\label{f:arc}
\end{figure}

\subsection{Photometric analysis}

The B and R frames were prepared using standard pre-reduction
techniques. Since there were only 2 frames per filter, cosmic rays
were removed by taking the lower pixel value in cases where a pixel in
one frame is significantly higher than the corresponding pixel in the
other frame. The photometric analysis was performed by means of the
SExtractor package (Bertin \& Arnouts, 1995) in the same way as Pierre
et al. (1997), but adapted to our data. The images were first slightly
smoothed to give the same PSF in B and R frames,
then the background was estimated using a 64 $\times$ 64 pixel
mesh. Source detections were claimed if at least 9 adjacent pixels
were above a threshold corresponding to 1.5 times the local noise
level. The CCD Sequence in M 92 (\cite{Christian}) observed during the
same run was used for photometric calibration. Stars VCS1, A, B
(probably variable) had to be removed because of obvious
inconsistencies.  Estimates of the photometric errors were taken
directly from the SExtractor analysis, and are less than $\sim 0.1$ for R$<22.5$ 
and less than $\sim 0.2$ for B$<23.5$.

The catalogue is estimated to be complete to R = 22.5 and B = 23.5. On
inspection of the detected objects above the completeness limit, we
found those objects with a SExtractor classification $<0.15$ may be
assumed to be galaxies, i.e., 275 objects. Changing the threshold does not
affect the outcome significantly because most of the galaxies are well
separated from stars (3/4 of the objects fall below 0.05 or above
0.95).

\begin{figure}
\resizebox{\hsize}{!}{\includegraphics{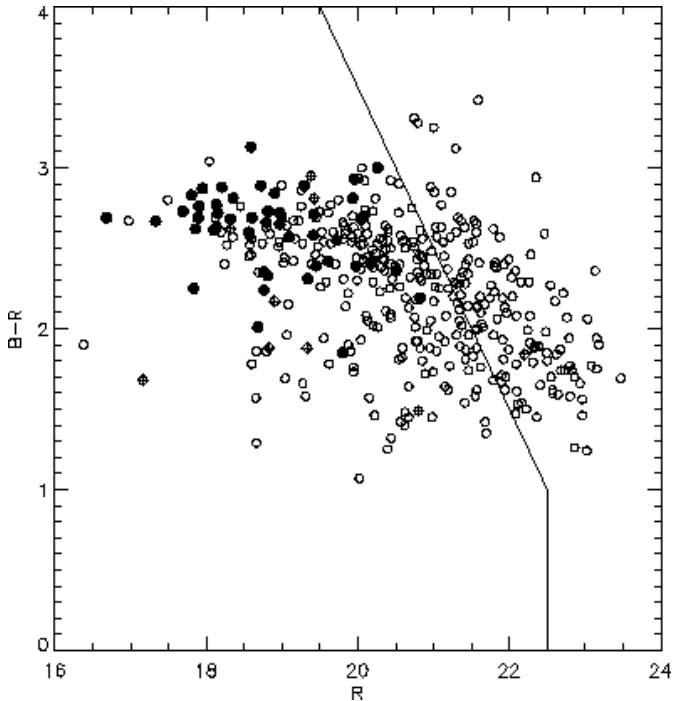}}
\caption []{ A colour magnitude diagram for all galaxies detected in
the R frame. The filled circles are for spectroscopically
confirmed cluster members, and the crosses are for the
non-members. The solid line gives the completeness limit.
\label{f:cm} }
\end{figure}

Fig.~\ref{f:cm} shows the colour magnitude diagram for all the
galaxies detected in the R-frame, and the corresponding magnitudes in 
the R and B bands were measured within the same apertures. The band of E/S0
sequence galaxies is discernable in Fig.~\ref{f:cm}; the
spectroscopically confirmed cluster members are shown to fall mostly
on the E/S0 sequence confirming that a large fraction of the
galaxies on the E/S0 sequence belongs to the cluster. The mean error in
B-R colour is $< 0.08$.

\subsection{Spectroscopy}

Grism O300 was used for the spectroscopy. It has a zero deviation at
5900~\AA, covers approximately 4700--7900~\AA, and gives a dispersion
of 3.59~\AA$/$pixel ($0.314\arcsec$). The slit has a width of
2\arcsec, i.e. 6.4 pixels, yielding a resolution of $\sim$23~\AA\
FWHM. Since there was only 1 frame per mask, cosmic rays were picked
out individually by eye and replaced by the median of the surrounding
pixels.  The internal Helium and Argon lamps was used for wavelength
calibration. The subsequent reduction was performed as described in
Pierre et al. (1997).  Redshifts were measured by a cross-correlation
method implemented in the MIDAS environment following Tonry and Davis
(1979). The cross-correlation results for each spectrum were checked
independently by eye.

The results from the cross-correlation analysis for all spectra are
presented in Table~\ref{spectro}.  Heliocentric correction has not
been applied, but is negligible at this resolution. The absolute error
in the velocity calibration is $\sim 200$\,km\,s$^{-1}$.

\begin{figure} 
\resizebox{\hsize}{!}{\includegraphics{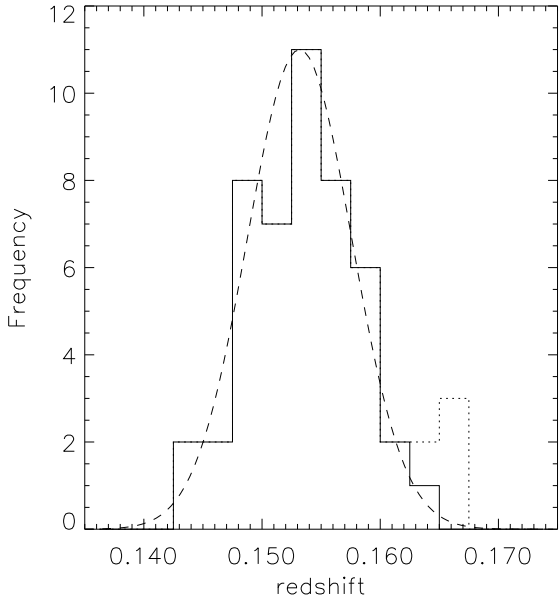}}
\caption []{ Cluster galaxy redshift histogram (bin size $\Delta z =
0.0025$). The galaxies considered to be cluster members are marked as
solid histogram while the dotted histogram are for the extended sample.
The dashed curve is a Gaussian with parameters corresponding
to the velocity distribution of the galaxies in the main sample.
\label{f:zhist} }
\end{figure}

As a first guess, galaxies are considered to be cluster members if
they lie within 3000 km\,s$^{-1}$ of the central cD galaxy, which
selects 47 (the main sample) out of the 60 galaxies. This procedure 
eliminates most of
the foreground and background galaxies without affecting the
dispersion measurements significantly.  If we relax the
velocity constraint and apply the usual $3\sigma-$clipping
technique then we have 51 cluster members (the extended sample). 
The cluster redshift
distribution for both samples is displayed in Fig.~\ref{f:zhist}. The histogram
includes all galaxies in the redshift range $z\sim 0.135 - 0.175$ in
Table~\ref{spectro} and a Gaussian corresponding to the velocity distribution 
of the main sample. The bi-weighted mean and scale for the main sample are 
$z=0.1532^{+0.0004}_{-0.0006}$ and
$\sigma=1148^{+190}_{-65}$\,km\,s$^{-1}$ correspondingly; and 
$z=0.1538^{+0.0009}_{-0.0006}$ and $\sigma=1401^{+160}_{-130}$\,km\,s$^{-1}$ for the
extended sample. It is
difficult to find an objective criterion for deciding which galaxies
are cluster members. Even with the sophisticated weighting scheme
employed by Carlberg et al. (1997), the determination of the weight
for each galaxy is still subjective. In Table~\ref{spectro}, we have
marked only the galaxies from the main sample as cluster members.

For the main sample we have enough redshifts to test whether or not
the galaxy velocities are drawn from a Gaussian distribution applying
various statistical tests for normality (e.g. D'Agostino \& Stephens
1986; {\small ROSTAT} -- Beers et al., 1990; Bird \& Beers, 1993). As
a result Anderson-Darling test (A$^2$) accepts the hypothesis for
normal distribution at 90\% significance, the combined
skewness and kurtosis test (B1 \& B2 omnibus test) at 97\% level and
the alternative shape estimators, asymmetry index and tail index based
on order statistics, were found to be $-0.21$ and $0.98$ respectively
which also show that the velocity distribution is drawn from a
Gaussian.

We can obtain a conservative estimate of the errors on the velocity
dispersion by comparing the dispersion from the extended and main
samples. When we take into account of the uncertainties in cluster
membership, a more conservative estimate of the errors should give the
velocity dispersion as $1200\pm 200$ km\,s$^{-1}$.

\begin{figure}
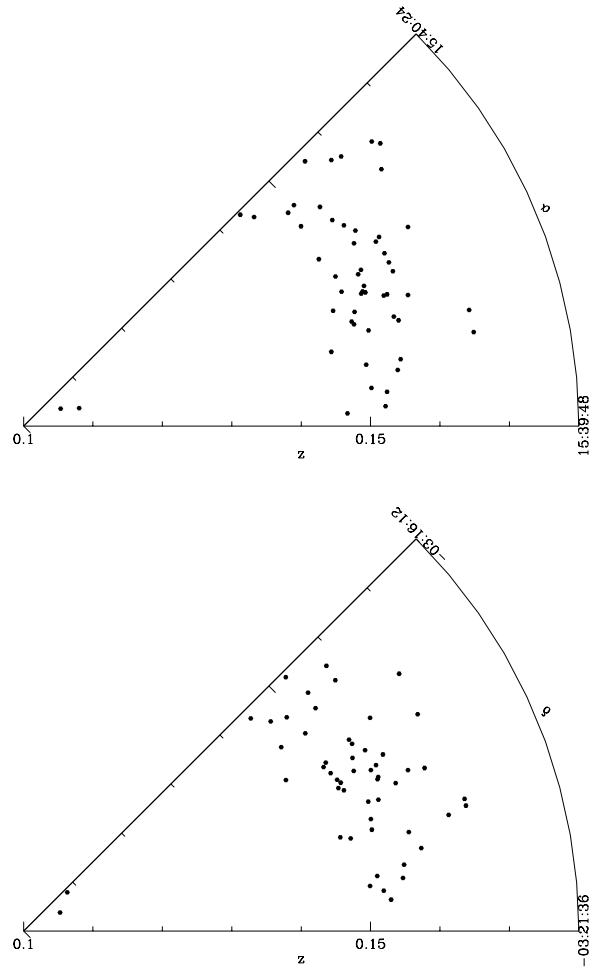
 
\resizebox{\hsize}{!}{\includegraphics{8809.f6a}}
\resizebox{\hsize}{!}{\includegraphics{8809.f6b}}
\caption[]{Wedge diagrams showing the distribution of galaxies
with measured redshifts.
\label{f:wedge} }
\end{figure}

We also investigated the presence of substructures in ($\alpha$,
$\delta$, $z$) space but no obvious signal was detected (see
Fig.~\ref{f:wedge}).  More redshifts are required for a proper
statistical analysis.

\section{X-ray Data}
We have observed the cluster with the {\it ROSAT} HRI and the 
{\it ASCA} GIS and SIS detectors. The HRI has a high spatial
resolution of $\sim 5\arcsec$, which provides a high resolution X-ray
surface brightness profile, but it has no energy resolution. {\it ASCA} on
the other hand has a low spatial resolution ($\sim 3\arcmin$) but
relatively high energy resolution and high sensitivity in the energy
range 1--10 keV, which provides a reliable gas temperature
measurement for clusters of galaxies. 

\subsection{Spectral analysis}
\begin{figure}
\resizebox{\hsize}{!}{\rotatebox{270}{\includegraphics*[115,44][556,726]{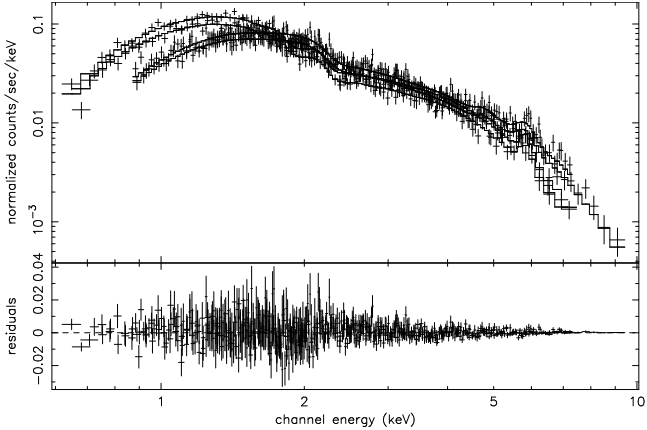}}}
\caption{{\it ASCA} spectra from the 4 detectors GIS2, GIS3, SIS0, SIS1. The
solid curves show the simultaneous fit to all 4 spectra using a
Raymond-Smith model with photoelectric absorption. The model fits also
take into account of the instrumental responses of the individual
detectors. \label{f:ascasp}}
\end{figure}

The cluster was observed with {\it ASCA} using both detectors of the Gas
Scintillation Imaging Spectrometers (GIS) and Solid-state Imaging
Spectrometers (SIS) in February 1996. The SIS detectors were operated
in 1-CCD mode.  The data was screened and cleaned according to the
standard procedures recommended (The ABC guide to {\it ASCA} data
reduction).  The spectra were extracted from the central $\sim
6.5\arcmin$ radius from the GIS2 and GIS3 detectors, excluding one
discrete source. Similarly, spectra were extracted from the central
$\sim 3\arcmin$ radius from the SIS0 and SIS1 detectors. A standard
blank-sky exposure screened and cleaned in the same way as the cluster
field was used for background subtraction by extracting a background
spectra from the same region on the detector as the cluster
spectra. The spectra were grouped into energy bins such that the
minimum number of counts before background subtraction was above 40,
which ensures that $\chi^{2}$ statistics would still be valid.  The 4 spectra
from each detector were simultaneously fitted with a Raymond-Smith
thermal spectra (Raymond \& Smith 1977) with photoelectric absorption
(Morrison \& McCammon 1983) from the {\small XSPEC} package
(Fig.~\ref{f:ascasp}). We adopted the abundance table with the relative
abundance of the various elements from Feldman (1992). The free
parameters were the gas temperature ($T_{g}$), Galactic neutral
hydrogen absorption column density (N(H)), metal abundance (abund) and
the emission integral. All 4 spectra were to have the same value for
the free parameters except for the emission integral, since the GIS
and SIS PSF were different and the extraction regions were
smaller for the SIS spectra compared to that of the GIS. The two GIS
spectra were assumed to have the same emission integral but different
from the SIS emission integrals. Results of the best simultaneous fit
to the 4 spectra along with fits to the individual spectra are
tabulated in Table~\ref{t:asca}. Only data in the energy range where
the effective area of the detectors are $> 10$ cm$^{2}$ were used for
the spectral fitting, i.e. 0.6--7.5 keV for SIS data and 0.85--10.0 keV
for GIS data.

\begin{table*}
\caption[]{Results on spectral fit to {\it ASCA} data}
\begin{flushleft}
\begin{tabular}{ccccccc}
\noalign{\smallskip}
\hline 
\noalign{\smallskip}
   & GIS & SIS & GIS+SIS & GIS* & SIS* & GIS+SIS* \\ \hline
kT$_{g}$ & $8.95^{+1.55}_{-1.24}$ & $7.33^{+0.97}_{-0.71}$ & $7.88^{+0.56}_{-0.52}$ & $10.51^{+1.21}_{-1.12}$ & $11.04^{+0.90}_{-1.21}$ & $10.36^{+0.64}_{-0.65}$ \\
abund & $0.21^{+0.11}_{0.12}$ & $0.22^{+0.11}_{0.10}$ & $0.32\pm0.08$
      & $0.20\pm0.13$ & $0.23\pm16$ & $0.22\pm0.07$ \\
N(H)  & $15.3\pm4.7$ & $22.5\pm2.2$ & $20.8\pm1.6$
      & 9.25 & 9.25 & 9.25 \\
$\chi^{2}$ & 0.55   & 0.75  & 0.91 
           & 0.56 & 0.90 & 0.95 \\ 
\noalign{\smallskip}
\hline
\end{tabular}
\end{flushleft}

{Notes:\\
 kT$_{g}$ - the gas temperature in keV;\\ 
 abund - the fractional solar metal abundance;\\ 
 N(H) - the neutral hydrogen column density in units of $10^{20}$ cm$^{2}$;\\ 
 $\chi^{2}$ - reduced $\chi^{2}$.\\ 
 col. 2 - fit to the combined GIS data;\\ 
 col. 3 - fit to the combined SIS data;\\ 
 col. 4 - simultaneous fit to GIS2, GIS3, SIS0 \& SIS1 spectra;\\ 
 col. 5,6,7 - same as col. 2,3,4 respectively, but N(H) was fixed to the 
 radio value and SIS data below 1 keV were not used.\\ 
 The quoted errors for each parameter correspond to the
90\% confidence range. \\
\label{t:asca}}
\end{table*}

The neutral hydrogen column density derived from the {\it ASCA} data were 2
times larger than the N(H) ($=9.25\times10^{20}$ cm$^{2}$) measured
from radio data by Starck (1992). If we try to fix N(H) to the value
determined by Starck (1992), then there is obvious discrepancy between
the model spectrum and the SIS data below 1 keV. Unfortunately, there
is no PSPC data available for this cluster to place definitive
constraints on the N(H) value. It is possible that there is a local
over-density of absorbing neutral gas along the line-of-sight to the
cluster, though it is more likely to be a calibration error for the
SIS detector. Calibration of the low-energy part of the SIS detector
is known to produce erroneous results such that it favours a high N(H)
inconsistent with PSPC results (Schindler et al. 1998 \& Liang et
al. 2000). In view of the possible calibration error for the SIS, the
data were also fitted with the above models with a fixed N(H) given by
Starck (1992) by excluding the SIS data below 1 keV.  The temperature
thus deduced was significantly higher than before. In the following
studies, we will adopt these parameters deduced from a simultaneous
fit of data from the GIS detectors in the energy range 0.85 to 10 keV
and the SIS detectors between 1 and 7.5 keV.

\subsection{{\it ROSAT} HRI data}
The cluster was observed by the {\it ROSAT} HRI in February (7.6ksec) and
August (36ksec) 1996. The X-ray centroid was found to be 15:40:08.1
$-$03:18:17, which is $\sim 1\arcsec$ from the position of the cD
galaxy 15:40:07.96 $-$03:18:16.7. The positional error for the X-ray
centroid is $\sim 5\arcsec$, hence the small apparent displacement
between the cD position and the X-ray centroid is insignificant.  The
X-ray surface brightness was obtained by extracting the photons in a
radius of $7\arcmin$ and the background was extracted from an annulus
of $8\arcmin-10\arcmin$ radius from the X-ray peak. Discrete X-ray
sources were excluded from the extraction. There were 8 discrete X-ray
sources in the HRI image.  Fig.~\ref{superover} shows the X-ray
contours overlaid on the optical image of the cluster field. The X-ray
image show significant substructure in the centre with an overall
elliptical appearance. The discrete X-ray sources at 15:40:07.2
$-$03:19:53 is embedded in the cluster emission.  The relative
astrometry between X-ray and optical was checked using 4 of the
discrete X-ray sources that had clear optical identification. The
X-ray positions had a maximum displacement of $\sim 2\arcsec$ relative
to the optical coordinates. The X-ray contours were adjusted to the
optical coordinate system using the 4 discrete X-ray sources, which
gave a relative astrometric accuracy of $\sim 0.5\arcsec$ between
optical and X-ray coordinates.

\begin{figure}
\resizebox{\hsize}{!}{\includegraphics{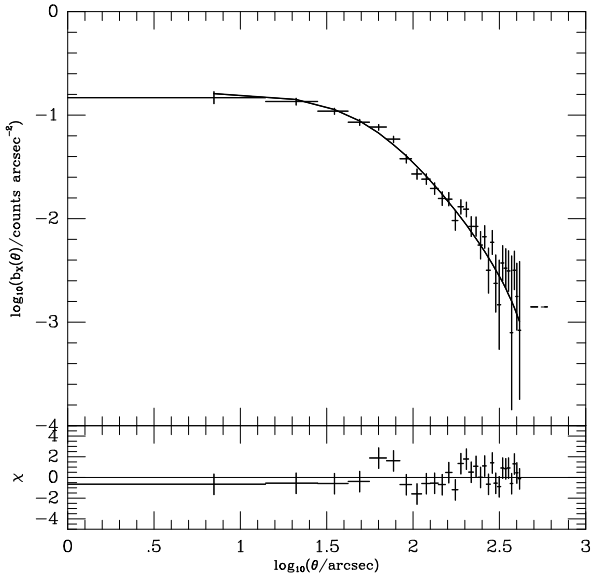}}
\caption{X-ray surface brightness from HRI data. The solid curve gives
the best $\beta$ profile after convolution with the HRI PSF. The
dashed horizontal line segment indicates the background level.
\label{f:beta}}
\end{figure}

A radial average of the X-ray surface brightness for the cluster is
shown in Fig.~\ref{f:beta}. A best fit $\beta$ profile (Cavaliere and
Fusco-Femiano 1976)
\begin{equation} 
S_{x}(r)=S_{0} [1+(\frac{r}{r_{0}})^{2}]^{-3\beta+1/2}
\label{eq:beta}
\end{equation}
convolved with the
instrument PSF is shown as a solid curve superimposed on the data.
The best fit gave $\beta=0.50^{+0.02}_{-0.03}$ and
$\theta=51^{+5}_{-6}\arcsec$.  The uncertainties quoted are
$1\sigma$. The total X-ray luminosity within the central $7\arcmin$
radius is $L_{x}\sim 8.95\times10^{44}$ ergs s$^{-1}$ in the {\it ROSAT}
band of 0.1--2.4 keV, assuming $N(H) = 9.25 \times10^{20}$ cm$^{2}$,
$kT_{g}= 10.4$ keV (or $1.2\times10^{8}$ K), and an abundance of
0.22. The X-ray luminosity thus deduced is consistent with that
estimated from the {\it ROSAT} all-sky survey (Pierre et al. 1997). The
corresponding bolometric X-ray luminosity is $L_{x}\sim
3\times10^{45}$ ergs s$^{-1}$. The central electron density was thus
derived to be $n_{e,0}\sim 5.92\times10^{-3}$ cm$^{-3}$. The central
cooling time for this cluster is $t_{cool} \sim 10^{10}$\,yr, greater
than a Hubble time.

\section{Analysis}
While the X-ray image show significant substructure in the cluster
indicating deviations from hydrostatic equilibrium, the cluster total
mass deduced from assumptions of dynamical equilibrium are still
reliable, as is shown by numerical simulations (Evrard et al 1996 \&
Schindler 1996).  Under the assumption of hydrostatic equilibrium and
spherical symmetry, the cluster total mass is directly related to the
intracluster gas properties as:
\begin{equation}
M_{tot}(r)=-\frac{rkT_{g}(r)}{\mu m_{p}G}(\frac{d\ln{n_{e}(r)}}{d\ln{r}}+\frac{d\ln{T_{g}(r)}}{d\ln{r}})
\label{eq:hydro}
\end{equation}

In general, a good fit can be found for the X-ray surface brightness
distribution using the parametrisation given in Eqn.~\ref{eq:beta},
which in turn gives the gas density as follows
if the gas is isothermal:
\begin{equation}
\frac{n_{e}(r)}{n_{e,0}}=[1+(r/r_{0})^{2}]^{-3\beta/2}
\label{eq:ne}
\end{equation}
Hence, the gravitational potential is given by
\begin{equation}
\phi(r)-\phi_{0}=\frac{3\sigma_{0}^{2}}{2}\ln{[1+(\frac{r}{r_{0}})^{2}]}
\label{eq:pot}
\end{equation}
where $\sigma_{0}^{2}\equiv\beta kT_{g}/\mu m_{p}$
and the total mass is given by
\begin{equation}
M_{tot}(r)=(\frac{3\sigma_{0}^{2}r_{0}}{G}) 
\frac{(r/r_{0})^{3}}{1+(r/r_{0})^{2}}
\label{eq:mtot}
\end{equation}

The lensing effects of the background galaxies by the cluster
gravitational field is directly related to the 2-D projection of the
total mass density. In this case, the projected total mass density is
given by
\begin{equation}
\Sigma^{2D}_{tot}(r)=\frac{3\sigma_{0}^{2}}{4Gr_{0}}\frac{2+(r/r_{0})^{2}}{[1+(r/r_{0})^{2}]^{3/2}}.
\label{eq:mass2d}
\end{equation}

If we consider galaxies as test particles in the cluster potential well, then 
Jean's equation for a collisionless, steady state, non-rotating spherically symmetric system gives
\begin{equation}
M_{tot}(r)=-\frac{r\sigma_{r}^{2}(r)}{G}(\frac{d\ln{n_{gal}(r)}}{d\ln{r}}+\frac{d\ln{\sigma^{2}_{r}(r)}}{d\ln{r}}+2\beta_{t})
\label{eq:jean}
\end{equation}
where $n_{gal}$ is the spatial galaxy number density, $\beta_{t}$ is
the anisotropy index and $\sigma_{r}$ is the radial velocity
dispersion.  The spatial galaxy number density is related to the
observed 2-D projection of the galaxy number density through the Abel
inversion given by
\begin{equation}
n_{gal}(r)=-\frac{1}{2\pi}\int_{r}^{\infty}\frac{d\Sigma^{2D}_{gal}(R)}{dR}\frac{dR}{\sqrt{R^{2}-r^{2}}},
\label{eq:proj}
\end{equation}
$\sigma_{r}$ and $\beta_{t}$ are related to the observed line-of-sight velocity dispersion $\sigma_{l}$ through
\begin{equation}
\Sigma^{2D}_{gal}(R) \sigma_{l}^{2}(R) = 2\int_{R}^{\infty} n_{gal}(r) \sigma_{r}^{2}(r)[1-\frac{R^{2}}{r^{2}} \beta_{t}] \frac{rdr}{\sqrt{r^{2}-R^{2}}}
\label{eq:vline}
\end{equation}

In the simple case, where the galaxy orbits are isotropic,
Eqn.~\ref{eq:jean} is equivalent to Eqn.~\ref{eq:hydro} with
$\sigma_{r}^{2}$ replaced by $kT_{g}/\mu m_{p}$.

If we make a further simplification by assuming that not only the gas but also the galaxies are isothermal, i.e. $\sigma_{r}(r)$ is a constant, then we have
\begin{equation}
\frac{n_{e}(r)}{n_{e,0}}=(\frac{n_{gal}(r)}{n_{gal,0}})^{\beta_{s}}
\end{equation}
where $\beta_{s}=\mu m_{p}\sigma_{r}^{2}/kT_{g}$.  Given the above
parametrisation for the X-ray surface brightness and the resultant
expression for $n_{e}$ given by Eqn.~\ref{eq:ne}, we deduce the
spatial galaxy density distribution as
\begin{equation}
\frac{n_{gal}(r)}{n_{gal,0}}=[1+(r/r_{0})^{2}]^{-\alpha}
\label{eq:3dgal}
\end{equation}
where $\alpha=3\beta/2\beta_{s}$.
The observed line-of-sight velocity dispersion is trivially given by 
$\sigma_{obs}=\sigma_{r}$ and $\alpha=3\sigma^{2}_{0}/2\sigma^{2}_{obs}$.

Alternatively, if we simplify the case by assuming that the galaxy
density distribution follows that of the total mass,
i.e. mass-follows-light, then from Jean's equation
(Eqn.~\ref{eq:jean}) we see that the galaxies can not be isothermal if
the gas is isothermal and the X-ray surface brightness is parametrised
as in Eqn~\ref{eq:beta}. The radial velocity dispersion is given by
\begin{equation}
\sigma^{2}_{r}(r)=\frac{1+(r/r_{0})^{2}/2}{1+(r/r_{0})^{2}/3}\sigma^{2}_{0}
\end{equation}
where again $\sigma^{2}_{0}\equiv\beta kT_{g}/\mu m_{p}$.
The line-of-sight velocity dispersion $\sigma_{l}$ can be deduced from Eqn.~\ref{eq:vline}.
However, the measured velocity dispersion is an average of $\sigma_{l}$ within a certain radius:
\begin{equation}
\sigma_{obs}^{2}(<R)=\frac{3}{4}[\frac{1+2(R/r_{0})^{2}-\sqrt{1+(R/r_{0})^{2}}}{(R/r_{0})^{2}}]\sigma^{2}_{0}.
\label{eq:obsdisp}
\end{equation}

In the case of \object{Abell~2104}, we have the observables
$S_{x}(R)$, $T_{g}$, $\Sigma_{gal}^{2D}(R)$, $\sigma_{obs}$. Since the
{\it ASCA} PSF was too poor to deduce a meaningful temperature profile, we
will assume that the gas is isothermal for the time being.  In the
following section we will study the cluster total mass deduced from
the various methods and examine their consistency using the simple
parametrised $\beta$-model given above.

\subsection{Mass estimate from optical data}
\begin{figure}
\resizebox{\hsize}{!}{\includegraphics{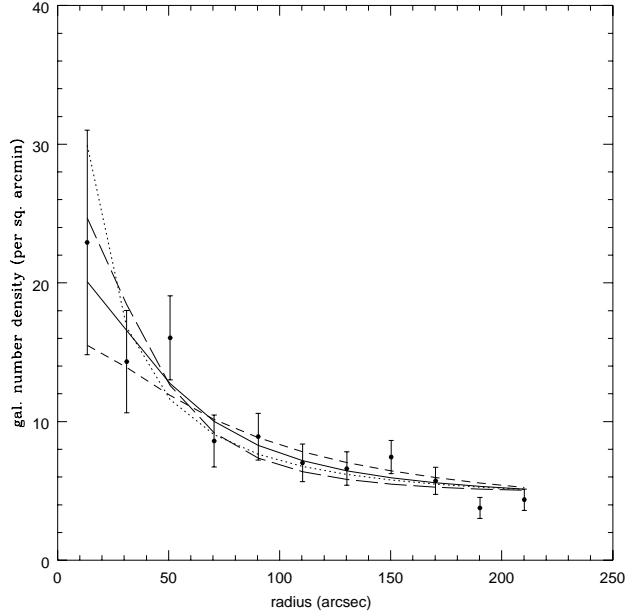}}
\caption{The radially averaged galaxy number density distribution
(with background). The curves show a number of
statistically consistent model fits to the data points. The 
short-dashed, solid and long-dashed curves corresponds to the $k=1/2,1,3/2$
cases of the family of curves given by Eqn.~\ref{eq:k}. The dotted
curve shows the 2D projection of the Navarro model (Navarro et al.
1996).
\label{f:2dgal}}
\end{figure}

The projected galaxy density distribution is consistent with a wide
range of models. The following family of
parametrised functions
\begin{equation}
\Sigma_{gal}(r)=\Sigma_{gal,0}[1+(r/r_{0})^{2}]^{-k}
\label{eq:k}
\end{equation}
were fitted to the projected galaxy density distribution after
background subtraction using the density of galaxies in the annulus
$220\arcsec$ to $240\arcsec$ as background. If we fix the core
radius to the X-ray determined value of $51\arcsec$, then we found the
best fit to be $k=1$ ($\chi^{2}=9.3$ with 10 degrees of freedom),
though $k=1/2$ to $3/2$ were also statistically consistent with the
observed data. Note that $k=1/2,1,3/2$ corresponds to spatial galaxy
distributions of the form given in Eqn.~\ref{eq:3dgal} with
$\alpha=1,3/2,2$ respectively.  The projected total mass density
distribution given by Eqn.~\ref{eq:mass2d} was also statistically
consistent with the projected galaxy density distribution
($\chi^{2}=9.66$ with 10 degrees of freedom), which means
mass-follows-light is not excluded.  The 2D projection of the
functional form $(r/r0)^{-1}[1+(r/r0)]^{-2}$ (Navarro et al. 
1996) was also found to be statistically consistent with the observed
galaxy distribution.  The projected galaxy distribution is shown in
Fig.~\ref{f:2dgal} along with the various model fits. The observed
galaxy density distribution is still declining towards the edge of the
image indicating a wider field is needed to reach the true ``edge'' of
the cluster. The X-ray data show that the cluster extends at least
out to a radius of $7\arcmin$ which is beyond the optical field of
view for the current observation. A wider field of view would help to
reject some of the above models.

If we estimate the total mass distribution from the galaxy density
distribution and velocity dispersion assuming that the galaxies are
isothermal, then $\sigma^{2}_{0}=(2\alpha/3)\sigma^{2}_{obs}$ where
the observed data give $\alpha \sim 1$ to $\sim 2$ and
$\sigma_{obs}=1200\pm 200$ km\,s$^{-1}$, implying that $\sigma_{0}
\sim 823$ to $\sim 1625$ km\,s$^{-1}$. Thus from Eqn.~\ref{eq:mtot}
the total mass is between $3.5\times10^{14} M_{\sun}$ and
$13.4\times10^{14} M_{\sun}$ within a radius of $220\arcsec$ (or
0.76\,Mpc), and between $6.8\times 10^{14} M_{\sun}$ and
$2.6\times10^{15} M_{\sun}$ extrapolating to $7\arcmin$ (or
1.46\,Mpc). Note that optical data alone does not constrain the mass
very well, even under assumptions such as isothermality of the galaxy
distribution and isotropy of the orbits.

On the other hand, if the galaxy distribution is not isothermal but
follows that of the mass then the measured velocity dispersion implies
that $\sigma_{0} \sim 1010\pm 165$ km\,s$^{-1}$ from
Eqn.~\ref{eq:obsdisp} and the total mass is $\sim
(10.2\pm3.7)\times10^{14} M_{\sun}$ within a radius of $7\arcmin$ (or
1.46 Mpc).

\subsection{Mass estimate from X-ray data}

The values of $\beta\sim 0.5^{+0.02}_{-0.03}$, $r_{0}\sim
51^{+5}_{-6}\arcsec$ and $T_{g}\sim 10.4\pm 0.6$ keV have been
determined from spatial analysis of the HRI data and the
spectro-analysis of the {\it ASCA} data respectively.  Thus from the X-ray
data, $\sigma_{0}^{2}\equiv \beta kT_{g}/\mu m_{p}$ implies a
$\sigma_{0}\sim 895\pm 45$ km\,s$^{-1}$ and a X-ray deduced total mass
of $\sim (8.0\pm 0.8)\times 10^{14} M_{\sun}$ out to a radius of
$7\arcmin$ (or 1.46 Mpc).

Note that if the galaxies are isothermal, then the X-ray deduced mass
is consistent with the optically deduced mass (or generalised
``Virial'' mass) if $\alpha \sim 1$. The X-ray and optical data are
also marginally consistent if mass-follows-light.

The total gas mass within $7\arcmin$ was found to be $\sim 7.8\times
10^{13} M_{\sun}$ which gives a gas fraction of $\sim 10$\% compared
to the X-ray deduced mass, but 5--10\% compared to the dynamically
deduced mass. The gas fraction within a radius of $r_{500}=1.14$\,Mpc
(where the over-density is 500 times the critical density of the
Universe) is $\sim 8$\%, which is lower than the average gas fraction
of $(20\pm 1.9)$\% for nearby hot ($kT_{g}>4$\,keV) non-cooling flow
clusters (Arnaud \& Evrard 1999).  The gas fraction within a radius of
1.46 Mpc gives a lower limit to the baryonic fraction. Since the
baryonic matter density predicted from the Big Bang nucleosynthesis
gives $\Omega_{b} \sim 0.04-0.06$ (Walker et al. 1991) from the
measured light element abundance, the lower limit of the baryonic
fraction of this cluster is thus consistent with
$\Omega_{M} \leq 1$.

\section{Discussions}
For the above simple models, we have shown that the X-ray deduced mass
is consistent with that from the optical data over the scale of 1--3
Mpc under the assumptions of dynamic equilibrium. In a recent paper by
Lewis et al. (1999), they also found the X-ray and dynamically deduced
mass were consistent for a sample of CNOC clusters at $z\sim0.3$.

On the other hand, in a study of a sample of clusters with giant arcs,
Allen (1998) found that the X-ray deduced mass was consistent with the
position of the giant arcs for cooling flow clusters but $2-3$ times
smaller than the lensing mass for non-cooling flow clusters. This was
then explained as a direct consequence of the theory that cooling flow
clusters were dynamically more relaxed than non-cooling flow clusters
since cluster mergers would certainly disrupt a cooling flow.  The
cooling time for \object{Abell~2104} is $t_{cooling} \sim 10^{10}$ yr
at the centre, thus there is no evidence for a cooling flow in this
cluster. Pierre et al. (1994) found a red tangential arc $7\farcs2$
from the centre of the cD galaxy (see Fig.~\ref{f:arc}). They found
that the projected mass within the arc to be $6\times 10^{12}
M_{\sun}$. Here we examine if the arc feature is consistent with the
simple cluster potential deduced from the X-ray data. Since the
projected density must reach the critical value at $7\farcs2$, it
requires $\sigma_{0} \sim 1380-1175$ km\,s$^{-1}$ for an arc redshift
in the range $z_{arc}\sim 0.5-3$ assuming the potential is spherically
symmetric. However, the X-ray data gave $\sigma_{0} \sim 895 \pm 45$
km s$^{-1}$ apparently inconsistent with the lensing deduced value,
indicating that in this very simplistic model the X-ray mass within
the arc radius appears to be $\sim 1.5-2$ times smaller than needed to
produce the giant arc. Our result appears to be consistent with the
results of Allen (1998). However, since the model we have adopted so
far is very simple and the arc radius is relatively small
($7\farcs2$), it is premature at this stage to suggest that the
lensing results are inconsistent with the X-ray data under the
assumptions of hydrostatic equilibrium and isothermal gas. As it was
pointed out in Pierre et al. (1994), the small arc radius is an
indication that the local cD potential is probably as important as the
global cluster potential in forming the arc feature. Indeed for most
clusters with giant arcs, the arc radii are barely larger than the
PSPC resolution and probably a few times larger than the HRI
resolution, hence an inconsistency between X-ray deduced mass from
simple models and that of the strong lensing deduced mass are not
sufficient to prove that the cluster is not in dynamic equilibrium. An
alternative explanation for the results of Allen (1998) could be that
the cooling flow clusters are well modelled by a cluster potential
similar to the type given by Eqn.~\ref{eq:pot}, but non-cooling flow
clusters have a different shape of gravitational potential, e.g. a
mass profile that has a broad component in the outer parts of the
cluster (e.g. Gioia et al. 1998). It would be difficult for the HRI to
reject a model of this kind since it has a high background level and
it would be easy to ``hide'' faint diffuse emission at large radii. In
our study of Abell~2104, the current optical image does not 
extend to the extent of the X-ray emission, thus we need wide-field
imaging to find out the true extent of the cluster. 

The X-ray emission in the centre of the cluster shows strong
ellipticity, the effect such asphericity has on the mass estimates
needs to be addressed since the mass estimates given above were
calculated under the assumption of spherical symmetry. Neumann \&
B\"ohringer (1997), estimated the effects of asphericity on mass
estimates of CL0016+16, and found that the total mass was only
changed by $\sim 2\%$ when the ellipticity was taken into account. The
ellipticity demonstrated in the X-ray image of \object{Abell~2104} is
no stronger than that of CL0016+16.

\begin{figure}
\resizebox{\hsize}{!}{\includegraphics{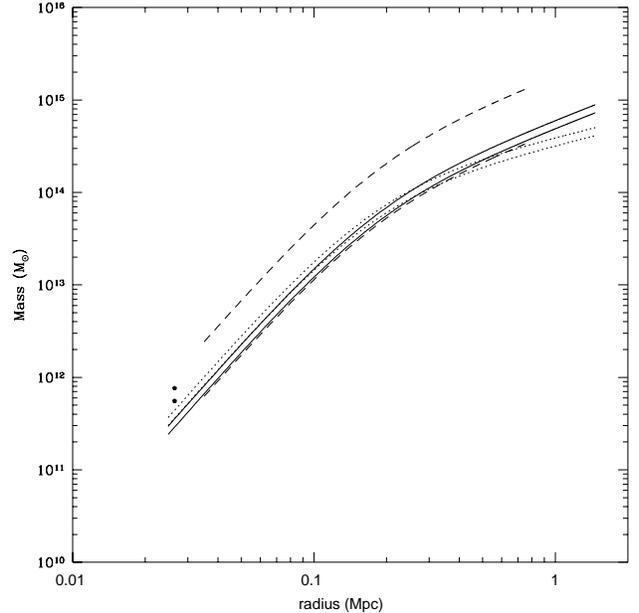}}
\caption{A comparison of 3D total mass distribution derived by the
various methods. The solid curves give the range of X-ray deduced mass
for isothermal gas; the dotted curves give the range of mass for the
polytropic model. The dashed curves give the range of dynamic mass
derived from the galaxy density distribution and velocity
dispersion. The curves are plotted only for regions were data is
available. The two stars show the range of mass estimates deduced from the
position of the lensing arc.
\label{f:mass}}
\end{figure}

So far we have only considered the isothermal gas models, but the
total mass given by Eqn.~\ref{eq:hydro} is more sensitive to $T_{g}$
than $n_{e}$. It is necessary to explore models with a temperature
gradient. Markevitch et al. (1998) found an almost universal decrease
in temperature in the outer regions over a radius of 0.3 to 1.8 Mpc in
a sample of 30 nearby clusters ($0.04<z<0.09$). They found that for a
typical 7 keV cluster, the observed temperature profile can be
approximated by a polytropic equation of state with $\gamma \sim
1.2-1.3$. If we assume that \object{Abell~2104} has a similar large scale
temperature profile, then we can quantify the mass ratio between the
polytropic and isothermal models as
\begin{equation}
\frac{M^{poly}_{tot}(r)}{M^{iso}_{tot}(r)}=\gamma (\frac{n_{e}(r)}{n_{e,0}})^{\gamma-1}.
\end{equation}
Since the X-ray emissivity has only a weak dependence on $T_{g}$ over
the 1--10 keV range (only a 10\% change), the X-ray surface brightness
varies insignificantly with $T_{g}$. We can then safely take the gas
distribution as determined from the isothermal case
(i.e. Eqn.~\ref{eq:ne}). Thus at $7\arcmin$ radius (1.46 Mpc), a model
with such a temperature gradient would give a mass that is $\sim 0.6$
times smaller than the isothermal case. This would cause the X-ray
deduced mass to be strongly inconsistent with the dynamically deduced
mass unless $\sigma_{r}^{2}$ increases with radius in a similar manner
as $T_{g}$. Note that a temperature profile that
decreases with the radius would also increase the total mass in the
inner cluster regions compared to the isothermal model, and thus
alleviate the discrepancy between the X-ray mass within the arc radius
and the position of the giant arc. Fig.~\ref{f:mass} shows the range
of mass profiles deduced from the various methods and models discussed
in the paper.

\section{Conclusions and Future prospects}
The rich cluster \object{Abell~2104} at a redshift of $z=0.1533$ was
found to have a high X-ray luminosity ($\sim 9.0\times 10^{44}$
ergs~s$^{-1}$ in [0.1-2.4] keV) and temperature ($10.4\pm0.6$ keV)
from {\it ROSAT} HRI and {\it ASCA} data. The central cooling time, $t_{cool} \sim
10^{10}$\,yr for this cluster indicates the absence of a cooling
flow. The galaxy velocity distribution showed that the cD galaxy was
at rest at the bottom of the cluster potential. The X-ray image shows
significant substructure in the centre of the cluster and an overall
elliptical appearance. It appears that the cluster has not yet reached
dynamical equilibrium.

As shown in Evrard et al (1996) and Schindler et al. (1996), the total
mass deduced from assumptions of dynamical equilibrium are not
significantly different from the true values. The total mass deduced
from the X-ray data assuming hydrostatic equilibrium is consistent
with the dynamic mass deduced from Jean's equation.  However, the
current data on the projected galaxy density distribution and our
knowledge of the galaxy orbits are limited for studies of cluster
dynamics, which allows a wide range of possible parametric functions
for the spatial galaxy density distribution without even attempting
the non-parametric methods of Merritt and Tremblay (1994) or
considering any anisotropic orbits. This can be improved by a deep
wide-field observation, to extend the galaxy number density
distribution to a large radius (up to 3 Mpc) and to allow a direct
measure of the cluster mass from a weak shear analysis. This would
allow us to definitively address the issue of whether or not the
cluster is in dynamical equilibrium and constrain the range of
possible total mass distributions allowed by the wide-range of data
from lensing effects to X-rays. In order not to bias the results and
incorporate a wide-range of the possible total mass density
distributions, a non-parametric method should also be employed. With
the launch of XMM and Chandra, we will soon able to obtain a
temperature profile and probe the X-ray emission at the edge of the
cluster which is crucial to the improvement of the X-ray mass
estimates.

\begin{acknowledgements} 
We would like to thank Emmanuel Bertin for his source extractor
program, Mark Birkinshaw for providing the convolution programs,
Monique Arnaud and J-L. Sauvageot for useful discussions on {\it ASCA}
data reduction, and Frazer Owen for providing the radio
image. T.C. Beers for providing the {\small ROSTAT} package. We
acknowledge the use of the {\small Karma} package
(http://www.atnf.csiro.au/karma) for the overlays.

\end{acknowledgements}

\begin{table*}
\caption[ ]{ Spectral analysis of \object{Abell~2104} }

\begin{flushleft}
\begin{tabular}{cccl@{\hspace{6mm}}lcllllc}
\noalign{\smallskip} \hline \hline \noalign{\smallskip} ID &
RA(J2000) & Dec(J2000) & \multicolumn{1}{c}{$z$} & 
\multicolumn{1}{c}{$\Delta V$ } & Q & \multicolumn{1}{c}{R} & 
\multicolumn{1}{c}{$\Delta$R} &
\multicolumn{1}{c}{B} & \multicolumn{1}{c}{$\Delta$B} & 
member \\
\hline \noalign{\smallskip}
102 & 15:39:53.0 & -03:18:45.0 &  0.1504(*) &  216 &  1 &  19.80 & 0.02 &  21.65
 &   0.03  &  Y \\
103 & 15:39:53.8 & -03:19:13.4 &  0.0068 &  178 &  2 &  20.80 & 0.03 &  22.29 & 
  0.03  &  N \\
105 & 15:39:56.1 & -03:18:36.7 &  0.1552 &  179 &  2 &  20.08 & 0.02 &  22.78 & 
  0.06  &  Y \\
106 & 15:39:57.5 & -03:19:41.9 &  0.1663 &  195 &  1 &  19.09 & 0.01 &  21.66 & 
  0.03  &  Y \\
107 & 15:39:58.9 & -03:17:20.0 &  0.1456 &  154 &  1 &  18.59 & 0.01 &  21.72 & 
  0.04  &  Y \\
108 & 15:39:59.7 & -03:19:35.8 &  0.1664 &  137 &  2 &  18.13 & 0.01 &  20.76 & 
  0.03 &  Y \\
109 & 15:40:00.6 & -03:18:34.2 &  0.1561 &  254 &  2 &  18.68 & 0.01 &  20.69 & 
  0.02  &  Y \\
110 & 15:40:02.2 & -03:17:23.3 &  0.1496 &  126 &  2 &  17.90 & 0.01 &  20.66 & 
  0.02  &  Y \\
111 & 15:40:03.1 & -03:20:11.0 &  0.1585 &  174 &  2 &  17.83 & 0.01 &  20.08 & 
  0.01  &  Y \\
112 & 15:40:04.0 & -03:18:46.8 &  0.1557 &  123 &  2 &  17.33 & 0.01 &  20.00 & 
  0.02  &  Y \\
113 & 15:40:05.4 & -03:19:27.1 &  0.1526 &  152 &  2 &  18.35 & 0.01 &  21.16 & 
  0.03  &  Y \\
114 & 15:40:06.4 & -03:18:19.8 &  0.1498 &  211 &  2 &  19.42 & 0.01 &  22.13 & 
  0.05  &  Y \\
115 & 15:40:07.9 & -03:18:15.8 &  0.1536 &  154 &  1 &  16.68 & 0.00 &  19.37 & 
  0.01  &  Y \\
116 & 15:40:08.5 & -03:18:06.1 &  0.1499 &  126 &  2 &  18.57 & 0.01 &  21.15 & 
  0.03  &  Y \\
117 & 15:40:10.2 & -03:18:33.5 &  0.0367(*) &  296 &  2 &  17.16 & 0.00 &  18.84
 &   0.01 &  N \\
118 & 15:40:11.2 & -03:17:56.4 &  0.1544 &  108 &  1 &  18.99 & 0.01 &  21.69 & 
  0.03 &  Y \\
119 & 15:40:12.4 & -03:18:48.6 &  0.1555 &  142 &  2 &  18.96 & 0.01 &  21.68 & 
  0.03 &  Y \\
120 & 15:40:13.7 & -03:18:02.2 &  0.1545 &  120 &  2 &  18.13 & 0.01 &  20.90 & 
  0.03 &  Y \\
123 & 15:40:18.7 & -03:17:28.0 &  0.1656 &  249 &  2 &  18.76 & 0.01 &  21.11 & 
  0.02 &  Y \\
124 & 15:40:19.4 & -03:18:08.3 &  0.1648 &  143 &  2 &  18.31 & 0.01 &  20.99 & 
  0.03 &  Y \\
125 & 15:40:20.7 & -03:17:48.1 &  0.1586 &  196 &  1 &  18.90 & 0.01 &  21.74 & 
  0.04 &  Y \\
127 & 15:40:23.3 & -03:18:52.9 &  0.2413 &  249 &  2 &  19.42 & 0.01 &  22.23 & 
  0.05 &  N \\
202 & 15:39:49.8 & -03:16:45.5 &  0.1467 &  232 &  1 &  18.72 & 0.01 &  21.61 & 
  0.04 &  Y \\
204 & 15:39:52.3 & -03:16:18.5 &  0.1526 &  130 &  1 &  17.80 & 0.01 &  20.63 & 
  0.02 &   Y \\
207 & 15:39:56.1 & -03:18:30.2 &  0.1502 &  259 &  2 &  20.08 & 0.02 &  22.78 & 
  0.06 &  Y \\
211 & 15:40:01.2 & -03:20:24.7 &  0.1557 &  143 &  1 &  20.20 & 0.02 &  22.22 & 
  0.04 &  Y \\
213 & 15:40:03.3 & -03:18:35.3 &  0.1505 &  102 &  2 &  20.19 & 0.02 &  22.61 & 
  0.05 &  Y \\
214 & 15:40:04.4 & -03:19:37.2 &  0.1476(*) &  219 &  2 &  18.97 & 0.01 &  21.62
 &   0.03 &  Y \\
215 & 15:40:05.2 & -03:19:39.0 &  0.1523 &  126 &  1 &  18.32 & 0.01 &  21.00 & 
  0.03 &   Y \\
216 & 15:40:05.9 & -03:19:07.7 &  0.1531 &  124 &  2 &  17.89 & 0.01 &  20.58 & 
  0.02 &  Y \\
217 & 15:40:07.3 & -03:18:59.8 &  0.1577 &  115 &  1 &  18.76 & 0.01 &  21.00 & 
  0.02 &  Y \\
218 & 15:40:08.3 & -03:18:20.5 &  0.1059 &  199 &  2 &  18.32 & 0.01 &  20.94 & 
  0.03 &  N \\
219 & 15:40:09.9 & -03:18:56.5 &  0.1624 &  159 &  1 &  18.56 & 0.01 &  21.16 & 
  0.02 &  Y \\
220 & 15:40:10.4 & -03:16:39.0 &  0.1580 &  110 &  1 &  17.95 & 0.01 &  20.82 & 
  0.02 &  Y \\
221 & 15:40:11.6 & -03:16:54.5 &  0.1489 &  106 &  2 &  19.41 & 0.01 &  21.99 & 
  0.04 & Y \\
223 & 15:40:16.6 & -03:18:09.4 &  0.1493 &   99 &  2 &  19.72 & 0.02 &  22.27 & 
  0.05 &  Y \\
224 & 15:40:19.1 & -03:19:41.9 &  0.1490 &  198 &  1 &  18.79 & 0.01 &  21.45 & 
  0.03 &  Y \\
225 & 15:40:20.3 & -03:18:52.2 &  0.1601 &  183 &  1 &  18.81 & 0.01 &  21.14 & 
  0.03 &  Y \\
227 & 15:40:21.8 & -03:16:25.3 &  0.1449 &  171 &  2 &  20.05 & 0.02 &  22.73 & 
  0.06 &  Y \\
228 & 15:40:23.5 & -03:18:00.4 &  0.1436 &  239 &  2 &  20.26 & 0.02 &  23.26 & 
  0.10 &   Y \\
302 & 15:39:50.5 & -03:20:49.9 &  0.1522 &  204 &  1 &  19.45 & 0.01 &  21.84 & 
  0.04 &   Y \\
304 & 15:39:52.4 & -03:19:33.2 &  0.0122 &  228 &  2 &  22.20 & 0.06 &  24.04 & 
  0.10 &   N \\
306 & 15:39:54.8 & -03:19:09.5 &  0.1545 &  132 &  1 &  18.60 & 0.01 &  21.29 & 
  0.03 &   Y \\
311 & 15:40:00.4 & -03:20:32.3 &  0.1516 &  196 &  2 &  19.61 & 0.02 &  22.03 & 
  0.04 &  Y \\
312 & 15:40:01.7 & -03:18:40.0 &  0.1498 &  122 &  2 &  18.15 & 0.01 &  20.87 & 
  0.02 &  Y \\
313 & 15:40:02.5 & -03:16:36.5 &  0.1084(*) &  195 &  2 &  18.90 & 0.01 &  21.07
 &   0.02 &   N \\
314 & 15:40:04.0 & -03:20:38.4 &  0.1552 &  115 &  2 &  17.86 & 0.01 &  20.48 & 
  0.02 &   Y \\
\noalign{\smallskip} \hline
\end{tabular}
\end{flushleft}
\end{table*}
\begin{table*}  
\begin{flushleft}
\begin{tabular}{cccl@{\hspace{6mm}}lcllllc}
\noalign{\smallskip} \hline \hline \noalign{\smallskip} ID &
RA(J2000) & Dec(J2000) & \multicolumn{1}{c}{$z$} & 
\multicolumn{1}{c}{$\Delta V$ } & Q & \multicolumn{1}{c}{R} & 
\multicolumn{1}{c}{$\Delta$R} &
\multicolumn{1}{c}{B} & \multicolumn{1}{c}{$\Delta$B} & 
member \\
\hline \noalign{\smallskip}
315 & 15:40:05.1 & -03:18:29.5 &  0.1529 &  228 &  1 &  18.81 & 0.01 &  21.54 & 
  0.03 &   Y \\
316 & 15:40:06.2 & -03:18:27.4 &  0.1577 &  216 &  2 &  19.95 & 0.02 &  22.88 & 
  0.07 &  Y \\
317 & 15:40:07.6 & -03:17:06.7 &  0.1530 &  161 &  1 &  19.29 & 0.01 &  22.18 & 
  0.05 &  Y \\
318 & 15:40:08.5 & -03:16:56.3 &  0.1577 &  132 &  1 &  18.20 & 0.01 &  21.08 & 
  0.03 &  Y \\
319 & 15:40:10.1 & -03:19:52.0 &  0.1573 & 152 &  1 &  19.34 & 0.01 &  21.65 &   0.03
 &  Y \\
320 & 15:40:11.4 & -03:20:46.7 &  0.2004(*) &  153 &  2 &  18.83 & 0.01 &  20.71
 &   0.01 &  N \\
321 & 15:40:12.0 & -03:20:21.1 &  0.0706 &  213 &  2 &  19.33 & 0.02 &  21.21 & 
  0.03 &   N \\
323 & 15:40:15.0 & -03:16:48.0 &  0.1535 &  190 &  2 &  19.93 & 0.02 &  22.74 & 
  0.06 &    Y \\
324 & 15:40:16.6 & -03:19:45.8 &  0.1635 &  154 &  1 &  19.97 & 0.02 &  22.36 & 
  0.05 &  Y \\
325 & 15:40:17.2 & -03:21:00.7 &  0.1531 &  177 &  2 &  17.69 & 0.01 &  20.42 & 
  0.02 &   Y \\
327 & 15:40:19.4 & -03:20:42.4 &  0.1503 &  173 &  1 &  18.10 & 0.01 &  20.71 & 
  0.03 &   Y \\
328 & 15:40:20.8 & -03:18:15.1 &  0.2849 &  181 &  2 &  19.38 & 0.01 &  22.33 & 
  0.06 &   N \\
329 & 15:40:22.6 & -03:18:14.8 &  0.1557 &  251 &  1 &  20.82 & 0.03 &  23.01 & 
  0.07 &   Y \\
\noalign{\smallskip} \hline \hline
\end{tabular}
\end{flushleft}
{Notes:\\
Column 1: internal reference number to Fig. \ref{chart}. \\
Column 2 \& 3: RA and Dec (J2000). Galaxy positions
were determined from the R image and should have an accuracy of $\sim$
0\farcs7 rms.\\
Column 4: redshift\\  (*) signifies the presence of emission lines: \\
\hspace*{5mm}102:  H$\beta$, H$\alpha$, [N\,{\sc ii}], [S\,{\sc ii}] \\
\hspace*{5mm}117:  H$\beta$, [O\,{\sc iii}], H$\alpha$, [N\,{\sc ii}], [S\,{\sc ii}] \\
\hspace*{5mm}214:  H$\beta$, H$\alpha$ \\
\hspace*{5mm}313:  H$\beta$, H$\alpha$, [N\,{\sc ii}], [S\,{\sc ii}] \\
\hspace*{5mm}320:  He\,{\sc i}, H$\alpha$, [N\,{\sc ii}], [S\,{\sc ii}] \\
\hspace*{5mm}214: [O\,{\sc ii}], [O\,{\sc iii}], H$\beta$
, H$\alpha$, [N\,{\sc ii}], [S\,{\sc ii}] \\
Column 5: $\Delta V$ is the internal measurement error and is related
to the correlation coefficient $C_{corr}$ by the formula $\Delta V =
k/(1+C_{corr})$ where $k \sim 870$ km\,s$^{-1}$ was determined by the Tonry and
Davis (1979) method. \\
Column 6: redshift measurement quality:\\
\hspace*{5mm}1: highest peak in the correlation function and checked by hand, \\
\hspace*{5mm}2: highest peak in the correlation function but unable to be 
checked by hand. \\
Column 7, 8, 9 \& 10: R, $\Delta$R, B and $\Delta$B magnitudes: \\
Column 11: Cluster member galaxy (within $\pm 3000$ km/s of the cD galaxy). \\

 \label{spectro}}
\end{table*}

\end{document}